\documentstyle{article}
\begin{document}
\baselineskip 17pt
\title{{\bf On the origin of multi-component anyon wave
functions }}
\author{ Ansar Fayyazuddin \\ \\  Institute
of Theoretical Physics \\
University of Stockholm \\ Vanadisv\"{a}gen 9 \\ S-113 46 Stockholm, Sweden}
\date{December 1992  }
\maketitle

\centerline{\bf Abstract}

\bigskip
\noindent
In this paper I discuss how the component structure of anyon wave functions
arises in theories with non-relativistic matter coupled to a Chern-Simons
gauge field on the torus.
It is shown that there exists a singular gauge transformation which
brings the Hamiltonian to free form.  The gauge transformation removes
a degree of freedom from the Hamiltonian.  This degree of freedom generates
only a finite dimensional Hilbert space and is responsible for the
component structure of free anyon wave functions.  This gives an understanding
of the need for multiple component anyon wave functions from the point of view
of Chern-Simons theory.

\noindent

USITP-92-15
\hfill
\newpage

\noindent
\section{Introduction}
Anyons, particles which obey fractional statistics \cite{lm,wil}
(for a review see e.g. \cite{rev}), are now a well established
phenomenon in theoretical physics.  A controversy over whether arbitrary
fractional statistics could be defined on compact surfaces has now been
resolved by the work of several independent groups \cite{einar,wen,lee}.
They find that anyon
wave functions with arbitrary rational statistics parameter
${\theta}=\pi q/p$ can be defined on the torus if the wave functions
have $p$ components and $N_{A}/\kappa$ is an integer (where
$N_{A}$ is the number of anyons).
In general one
cannot simultaneously diagonalize the operators which translate the anyons
along
the different cycles of the surface.  However, one can pick a basis of wave
functions so
that the component index indicates the phase it picks up under translation
along one
cycle while translation along the other cycle shifts the component index by one
unit.

	In this paper I reconsider the problem of non-relativistic
particles coupled to a U(1) Chern-Simons gauge field on the torus.  I
will consider a non-relativistic quantum field theory of bosons minimally
coupled to the gauge field.
The structure of the argument is as follows.  Section two deals with
the gauss constraint imposed by the Chern-Simons term and its solution
on the torus.  In section three the physical degrees of freedom of the gauge
field are quantized followed by a discussion of the particle vacuum
which is essentially a summary of an argument due to Polychronakos \cite{poly}.
In section four the first quantized Hamiltonian is derived and the center
of mass Hamiltonian is explicitly diagonalized.  It is demonstrated in section
five that only single component wave functions are needed in the full theory.
Finally in section six it is shown that when the Hamiltonian is brought
to free form by a singular gauge transformation the single component
wave functions in the original theory
become multiple component wave functions.  It is argued that this is a
consequence of removing a dynamical degree of freedom from the
center of mass Hamiltonian
which only generates a $p$ dimensional Hilbert space.

Anyons on tori have been the subject of a number of recent investigations.
In particular, discussions addressing similar issues to the
ones presented in sections two
and the first part of section four can be found in \cite{lee,il,ho,lech}
As mentioned earlier the discussion of the particle vacuum in section four
is essentially a summary of arguments originally presented in \cite{poly}.

Throughout this paper the torus will be taken to be the $L_{1} \times L_{2}$
rectangle in the $xy$ plane with opposite sides identified and the
Chern-Simons coupling will be given by $\kappa = p/q$, where $p$ and $q$ are
relatively prime integers.
The modular parameter $\tau$ appearing in the Jacobi theta
functions is given by $\tau = iL_{2}/L_{1}$.

\section{Solving the Chern-Simons constraint}

This section will deal with the problem of
solving the Gauss Law constraint imposed by the Chern-Simons term in the
Lagrangian
and isolating the remaining degrees of freedom of the gauge field which
are to be quantized.
  On the torus the Gauss Law constraint does not completely determine the
gauge potential, unlike the situation on the plane.
This is because the Gauss Law only constrains
the curvature associated with the C-S connection and does not completely
specify
the connection (up to gauge transformations).  In fact, it is well known
that to
specify the connection completely on topologically non-trivial spaces one has
to
specify the Wilson lines $\exp i{\oint a_\mu}dx^\mu$ along all non-trivial
loops \cite{wy}.  In mathematical
language this amounts to specifying the cohomology class of the connection.  On
the
plane the de Rham cohomology group is trivial so that the curvature {\em does}
determine the
connection up to gauge transformation.  This is not the case on the torus.

There is no entirely natural way of dividing the gauge field into real
and constrained degrees of freedom.  In particular the Wilson loops around the
non-trivial cycles of the torus are affected by the amount of flux coming out
of the
torus and therefore depend implicitly on the constraint.
To see that this is the case one can convince oneself that the
flux flowing out of an area bounded by two loops along, say, the x-axis is
given by the product of the Wilson loops evaluated along these two (oppositely
oriented)
loops.  Therefore the Wilson loops must know about not only the flux
flowing through the holes of the
torus but also the flux flowing out of the torus.
Nonetheless it is possible to divide the gauge field into real and
constrained degrees of freedom by specifying a canonical solution to the
constraint so that the connection is unambiguously solved for.  Any terms
which one can add to the connection while preserving the
constraint will be real degrees of freedom and should be quantized.
The remaining part of
this section will be a concrete elaboration of this point.

Consider the following Lagrangian:
\begin{eqnarray}
{\cal L} &=& \frac{\kappa}{4\pi} {\epsilon}^{\mu\nu\rho}a_{\mu}
{\partial}_{\nu}a_{\rho} + {\psi}^{\dag}iD_{0}\psi
- \frac{1}{2m}{\left(D_{j}\psi\right)}^{\dag}{\left(D_{j}\psi\right)}
\nonumber \\
D_{0} &=& {\partial_0} - i{a_0}   \nonumber \\
D_{j} &=& {\partial_j} - i{a_j}  \nonumber \\
\end{eqnarray}
Where $a_\mu$ is the Chern-Simons gauge field
and the fields $\psi$ are bosonic matter fields.  From the above Lagrangian we
get the constraint:
\begin{equation}
0= \frac{\delta {\cal L}}{\delta a_0} = \frac{\kappa}{2\pi}f_{12} + J_{0}
\end{equation}
where
\begin{equation}
J_{0} = {\psi}^{\dag}{\psi}
\end{equation}
It is important to realize that if the gauge field supports a non-zero total
flux then there will be no globally well defined gauge potential on the torus.
The
mathematical reason for this is that the U(1) bundle over the torus is
twisted.  This can be seen by trying to calculate the
total flux flowing out of the torus:
\begin{eqnarray}
\phi & = & \int\int dxdy \left({\partial_x}{a_y}-{\partial_y}{a_x}\right)
\nonumber \\
& = & \int_{y_0}^{{y_0}+L_2}dy\left({a_y}\left({x_0}+{L_1},y\right)-
{a_y}\left({x_0},y\right) \right) \\
& - & \int_{x_0}^{{x_0}+L_1}dx\left({a_x}\left(x,{y_0}+{L_2}\right)
-{a_x}\left(x,{y_0}
\right)\right)
\end{eqnarray}
If the gauge field were periodic along the two cycles of the torus the
flux would vanish.  If the gauge field does support flux and satisfies
the following "quasi-periodic" boundary conditions:
\begin{eqnarray}
{a_j}\left({\vec r}+{\hat e}_{i}L_{i}\right) = {a_j}\left({\vec r}\right) +
\partial_{j}\Lambda_i
\end{eqnarray}
where the $\Lambda_i$ are such that $\exp i\Lambda_{i}\left(x,y\right)$ are
well defined
(i.e. single-valued) gauge transformation on the torus \footnote{When there is
an external
electromagnetic field present the condition that $\exp
i\Lambda_{i}\left(x,y\right)$ be
single valued can be relaxed to requiring that $\exp
i\left(\Lambda_{i}^{cs}\left(x,y\right)+
\Lambda_{i}^{ext}\left(x,y\right)\right)$ be single valued on the torus where
$\Lambda_{i}^{ext}\left(x,y\right)$ is the corresponding gauge function of the
external
field.}.  and if, in addition, the matter fields satisfy:
\begin{eqnarray}
\psi\left({\vec r}+{\hat e}_{i}L_{i}\right) =
\exp\left(i{\Lambda_i}\right)\psi\left({\vec r}\right)
\end{eqnarray}
then all gauge invariant observables such as the $J_{\mu}$ are well defined
on the torus and there is no problem with consistency.  I will work in
the $\nabla\cdot{\bf a}=0$ gauge, in which the Gauss law constraint can be
written as:
\begin{eqnarray}
\nabla^{2}a_{i}= \frac{2\pi}{\kappa}\epsilon^{ij}{\partial_j}J_0
\end{eqnarray}
It remains to give a canonical solution to the above equation and to
identify the remaining
degrees of freedom.  I pick the following solution:
\begin{eqnarray}
\hat{a}_{1} & = & \frac{2\pi}{\kappa}\left(\frac{yQ}{L_{1}L_{2}}
- \int {\partial_2}G\left(
\vec{r\prime}-\vec{r}\right)J_{0}d^{2}r\prime \right) \nonumber \\
\hat{a}_{2} & = & \frac{2\pi}{\kappa}\int {\partial_1}G\left(
\vec{r\prime}-\vec{r}\right)J_{0}d^{2}r\prime
\end{eqnarray}
where
\begin{equation}
Q= \int d^{2}r J_0
\end{equation}
is the particle number operator.
$G$ is the periodic Green's function on the torus:
\begin{eqnarray}
\nabla^{2}G\left(r\right) = \delta\left(\vec{r}\right) - \frac{1}{L_{1}L_{2}}
\end{eqnarray}
and is given by \cite{il,salam}:
\begin{eqnarray}
G\left(x,y\right) = \frac{1}{4\pi}ln \frac{{\mid \theta_1\left(z\mid\tau\right)
\mid}^{2}}
{{\mid {\theta_1}\prime \left(0\mid\tau\right) \mid}^{2}} +
\frac{y^2}{2L_{2}L_{1}}
\end{eqnarray}
where $z=x+iy$ and $\theta_1$ is the odd Jacobi theta function.
The above solution fixes $\Lambda_{1}=0$ and $\Lambda_{2}=2\pi Qx /\kappa L_1$.
It is easy to see that the only terms consistent with the constraint
and the choice of transition functions (the $\Lambda_{i}$s) that
one can add to $a_{i}$ are position independent terms:
\begin{equation}
a_{i}= \frac{\theta_i}{L_i}+\hat{a}_i
\end{equation}
In fact, only $\theta_{i}$ mod $2\pi$ is observable
(the rest being gauge equivalent to 0)\footnote{It should be clear
from the context
whether the $\theta_i$ stand for the Jacobi theta functions or the gauge
degrees of freedom.}.
The single valued transition functions
respect the seperation of the $\theta_{i}$ from the
${\hat a}_{i}$.

The flux flowing out of the torus ($2\pi Q/\kappa$) must be quantized
according to the Dirac quantization condition which follows from requiring
that the holonomy of any homotopically trivial closed path should be
well defined.  This imposes the condition that $Q/\kappa$ be an integer.
So the theory restricts the number of particles to be an integer multiple of
$p$.  This condition also makes the transition functions single valued on
the torus.

\section{Quantization and the Structure of the Vacuum}

I turn now to the quantization of the $\theta_{i}$.  The relevant term
in the Lagrangian is:
\begin{eqnarray}
\frac{\kappa}{4\pi}\epsilon^{ij}\int\dot{a}_{i}a_{j}d^{\left( 2 \right)}r
=\frac{\kappa}{4\pi}\left(\dot{\theta_{1}}\theta_{2}-\dot{\theta_{2}}
\theta_{1}
-\frac{\pi}{\kappa}\left({\theta_2}\dot{Q}-\dot{\theta_{2}}Q\right) +
\int\left(\dot{\hat a_1}\hat{a_2}-\dot{\hat a_2}\hat{a_1}\right)\right)
\end{eqnarray}
By partial integration of the action one can remove all the time derivatives
from $\theta_2$ up to a total derivative term which contributes a
surface term irrelevant to quantization.
The variation of the action with respect to $\dot{\theta_{1}}$ gives the
corresponding
conjugate momentum \footnote{One may of course treat both $\theta_{1}$
and $\theta_2$ as
coordinates but then one has to use Dirac brackets to quantize since the
momentum conjugate to $\theta_2$ vanishes identically.  Of course both
procedures give the same final result.  I thank T. H. Hansson for pointing
out an error in an earlier version of the manuscript concerning this point
and bringing ref \cite{fj} to my attention where this question is discussed in
its generality}:
\begin{eqnarray}
\Pi \equiv  \frac{\delta S}{\delta\dot{\theta_{1}}} =  \frac{\kappa}{2\pi}
\theta_{2}
\end{eqnarray}
Imposing the canonical commutation relations gives:
\begin{eqnarray}
\left[{\theta_1},\Pi\right] = i =
\frac{\kappa}{2\pi}\left[{\theta_1},{\theta_2}\right]
\end{eqnarray}

Finally, it is necessary to construct a vacuum on which physical states
can be built.
The first quantized Hamiltonian will depend on the particle coordinates,
momenta and the phases $\theta_i$ which are really global degrees of freedom
(they are to be interpreted as the amount of flux flowing through the holes
of the torus).  A basis for the Hilbert space is provided by
the set of states \{$\mid \theta_{1}\rangle\mid r_{1}\ldots r_{n} \rangle$\}.

Now try to construct a complete set of states in the $\theta$ sector.
A natural first guess is to construct states in the '$\theta_1$
representation': \{$\mid {\theta_{1}} \rangle$\},
$\hat{\theta_1} \mid {\theta_1} \rangle = {\theta_1}\mid {\theta_1} \rangle$,
$\hat{\theta_2} \mid {\theta_1} \rangle =
\frac{-i2\pi}{\kappa}\frac{\partial}{\partial{\theta_1}} \mid {\theta_1}
\rangle$ with the completeness relation: $\int d\theta_{1} \mid {\theta_1}
\rangle
\langle \theta_{1} \mid = 1$.  Since the $\theta_i$ are not observables the
summation overcounts by including physically equivalent states, states related
by gauge transformations.  The only
observables which can be constructed from the $\theta_i$ are
$U_{i}=\exp i\theta_i$
and therefore it is reasonable to construct a complete set of states with
respect to these observables.  The $U_i$ satisfy
\begin{eqnarray}
U_{1}U_{2} & = & U_{2}U_{1}\exp{i\frac{2\pi}{\kappa}}  \\
U_{1}{U_2}^{p} & = &{U_2}^{p}U_{1}
\end{eqnarray}
Let $\mid \theta_{1}, \alpha_{2}\rangle$ represent a state on which
\begin{eqnarray}
U_{1}\mid \theta_{1}, \alpha_{2}\rangle & = & e^{i\theta_{1}}\mid \theta_{1},
\alpha_{2}\rangle \\
{U_2}^{p}\mid \theta_{1}, \alpha_{2}\rangle & = &
e^{ip\alpha_{2}}\mid \theta_{1}, \alpha_{2}\rangle
\end{eqnarray}
The new completeness
relation reads $\int_{0}^{2\pi} \int_{0}^{2\pi} d\alpha_{1} d\alpha_{2} \mid
\alpha_{1}, \alpha_{2} \rangle \langle \alpha_{1},\alpha_{2} \mid = 1$.
Even this relation is not quite correct since the state
$\mid \theta_{1}+2\pi , \alpha_{2} \rangle$ does not lie on the same ray
as $\mid \theta_{1} , \alpha_{2} \rangle$ \cite{poly} even though
they represent the
same physical state.  This is easily seen from the fact that the operators
which perform these gauge transformations,
$T_{i} = \exp\left(-2\pi \frac{\partial}
{\partial \theta_{i}}\right)$, can be represented as
$T_{2}= \exp i\kappa\theta_1$ and $T_{1} = \exp -i\kappa\theta_2$
and satisfy
$T_{1}T_{2} = T_{2}T_{1}e^{-i2\pi\kappa}$.  The irreducible representaions
of the $T_i$ are q dimensional.  The states {$\mid \theta_{1} +
2\pi l,\alpha_{2} \rangle$}, $l=0,...,q-1$ are eigenstates of $T_2$ but
transform into each other under
the action of $T_1$ and do not lie on the same ray.  Thus the full Hilbert
space consists of q copies of the physical Hilbert space.  Since
the $\theta_i$ are being treated as phases, each physical sector of the Hilbert
space just specifies a way of picking a branch for the phases and one is
free to remain within one such sector.  Indeed all
physical observables commute with these transformations and thus restriction
to one sector is equivalent to fixing a gauge.

Having constructed a basis of states in the $\theta_1$ representation
in the gauge field sector of the theory any eigenvector of
$\theta_2$ can be expressed as a linear superposition of these states.
Which states does one need in order to construct the eigenstate of
$\exp i\theta_2$ with eigenvalue $\exp i\delta$?
To specify the state unambiguously one has to give the value of
$\exp {ip\theta_{1}}$ as well.  Consider the action of
$U_2$ on the state $\mid \theta_{1}, \beta \rangle$.  Since $\theta_{2} =
-i\frac{2\pi}{\kappa}\frac{\partial}{\partial\theta_1}$, the state
$\mid \theta_{1}, \beta \rangle$ is mapped on to
$\mid \theta_{1}-\frac{2\pi}{\kappa}, \beta \rangle$ upto a phase factor.
After p actions of $U_2$ the state returns to itself up to a phase
(recalling
that the states $\mid \theta_{1}+2\pi q, \beta \rangle$ and $\mid \theta_{1}
, \beta \rangle$ lie on the same ray).  Therefore $\exp i\theta_2$ can
always be diagonalized by the $p$ states
\{$\mid \alpha_{1}+2\pi l/\kappa, \beta \rangle$\},
$l = 0 \ldots p-1$, with possible eigenvalues $\beta + 2\pi n/\kappa$
where $\exp ip\alpha_1$ is the eigenvalue of $U_{1}^{p}$ \cite{poly}.
If $\delta$ belongs to the set \{$\beta + 2\pi n/\kappa$\} then
the corresponding eigenstate is a linear combination of the
$p$ states \{$\mid \alpha_{1}+2\pi l/\kappa, \beta \rangle$\}.
Thus the gauge field Hilbert space is divided
up into sectors
labeled by the eigenvalues $U_{1}^{p}=e^{ip\alpha_1}$
and ${U_2}^{p} = e^{ip\alpha_2}$ and each sector is
$p$ dimensional \cite{poly}.
With respect to the operators $U_1$ and $U_2$ the physical Hilbert space
has the form ${\cal H} = \stackrel{\bigoplus}
{\alpha_{1},\alpha_{2}} {\cal H}_{\alpha_{1}, \alpha_{2}}$ where
each ${\cal H}_{\alpha_{1}, \alpha_{2}}$ is $p$-dimensional.
The particle vacuum state is then $p$-dimensional.
This direct sum structure of
the Hilbert space will turn out to be intimately related to the multi-component
structure of anyon wave functions.

\section{The Schr\"{o}dinger Equation}
In this section I will write down the Hamiltonian in first quantized form
and define the Schr\"{o}dinger wave functions.  I will point out some new
qualitative features in the Hamiltonian which distinguish the torus from
the plane.  In particular I will concentrate on the center of mass
Hamiltonian and argue that it contains essentially all the qualitatively
new features of the Hamiltonian on the torus.

Following Jackiw and Pi \cite{jp}, I take the Hamiltonian to be:

\begin{eqnarray}
{\cal H}= -\frac{1}{2m} {\left(D_{i}\psi \right)}^{\dag}D_{i}\psi
\end{eqnarray}
where the covariant derivatives have already been defined above.
The wave functions are defined by:
\begin{eqnarray}
\phi\left(\theta ,x_{i},y_{i} \right) = \langle \theta \mid
\psi\left(r_{1}\right)\ldots \psi\left(r_{n}\right) \mid \phi \rangle
\end{eqnarray}
To make the notation less cumbersome I have adopted the follwing abbreviated
notation: $\theta$ stands for $\theta_1$,
the eigenvalue of ${U_2}^p$ issuppressed, and $\mid \theta \rangle$
is an abbreviation of $\mid \theta \rangle \mid 0 \rangle $, $\mid 0 \rangle$
being the particle vacuum.

The Schr\"{o}dinger equation is then given by:

\begin{eqnarray}
i\frac{\partial}{\partial t} \phi\left(\theta ,x_{i},y_{i} \right) & = &
\langle \theta \mid
\left[ \psi\left(r_{1}\right)\ldots \psi\left(r_{n}\right), H\right] \mid \phi
\rangle
\end{eqnarray}
This determines the first quantized Hamiltonian, in the $N_A$ particle
sector, to be:
\begin{eqnarray}
H & = & -\frac{1}{2m} \sum_{\alpha=1}^{N_A}{\vec D_\alpha}\cdot{\vec D_\alpha}
\\
{\vec D}_{\alpha} & = & {\vec \nabla}_{\alpha} - i{\vec a}_{\alpha} \\
a_{\alpha x} & = & \frac{\theta}{L_1} + \frac{2\pi N_A}{\kappa L_{1}L_{2}}
y_{\alpha}
+ \frac{2\pi}{\kappa}\sum_{\beta\neq \alpha} \left(
\frac{\partial}{\partial {y_\alpha}}G\left(x_{\alpha}-x_{\beta},y_{\alpha}-
y_{\beta}\right) \right) \\
a_{\alpha y} & = & -i\frac{2\pi}{\kappa}\frac{\partial}{\partial\theta} -
\frac{2\pi}{\kappa}\sum_{\beta\neq \alpha} \left(
\frac{\partial}{\partial {x_\alpha}}G\left(x_{\alpha}-x_{\beta},
y_{\alpha}-y_{\beta}\right) \right)
\end{eqnarray}

where $G$ is the periodic Green's function given above.  The expressions
for the ${\vec a_\alpha}$ can be written in the simpler form:
\begin{eqnarray}
a_{\alpha x} & = & \frac{\theta}{L_1} - \frac{2\pi N_A}{\kappa L_{1}L_{2}}Y
	+\frac{i}{2\kappa} \frac{\partial}{\partial x_{\alpha}}
	\sum_{\beta \neq \alpha} ln \frac{\theta_{1}^{\ast}\left(z_{\alpha}-
z_{\beta} \mid\tau
\right)}{\theta_{1}\left(z_{\alpha}-z_{\beta} \mid\tau \right)} \\
a_{\alpha y} & = & -i\frac{2\pi}{\kappa} \frac{\partial}{\partial \theta}
+\frac{i}{2\kappa} \frac{\partial}{\partial y_{\alpha}}
\sum_{\beta\neq \alpha} ln \frac{\theta_{1}^{\ast}\left(z_{\alpha}-z_{\beta}
\mid\tau
\right)}{\theta_{1}\left(z_{\alpha}-z_{\beta} \mid\tau \right)}
\end{eqnarray}
where $X$ and $Y$ are the center of mass coordinates defined by
$X \equiv \frac{1}{N_A}\sum_{\alpha =1}^{N_A}x_{\alpha}$ and
$Y \equiv \frac{1}{N_A}\sum_{\alpha =1}^{N_A}y_{\alpha}$.

The Hamiltonian conveniently splits up into a center of mass plus a relative
piece which commute with each other.  The wave functions will then be of the
product form $\psi^{cm}\otimes\psi^{rel}$ where each factor will satisfy
the Schr\"{o}dinger equation with respect to the appropriate Hamiltonian.
It is illuminating to see the explicit form of the Hamiltonians:
\begin{eqnarray}
H^{cm} & = & -\frac{1}{2mN_A} \left[ \left(\frac{\partial}{\partial X}
-i\frac{N_{A}\theta}{L_1} -i \frac{2\pi {N_A}^{2}}{\kappa L_{1}L_{2}}Y
\right)^{2}
+ \left( \frac{\partial}{\partial Y}-\frac{2\pi N_{A}}{\kappa L_2}
\frac{\partial}{\partial \theta} \right)^{2} \right] \\
H^{rel} & = & -\frac{1}{2mN_A}\sum_{\alpha,\beta, \alpha\neq \beta}
\left[ \left(\frac{\partial}{\partial x_\alpha} - \frac{\partial}{\partial
x_\beta} -
\frac{1}{2\kappa} \left( \frac{\partial}{\partial x_\alpha} - \frac{\partial}
{\partial x_\beta} \right)\Lambda
\right)^{2} \right. \nonumber \\
& + & \left. \left(
\frac{\partial}{\partial y_\alpha} - \frac{\partial}{\partial y_\beta} -
\frac{1}{2\kappa} \left( \frac{\partial}{\partial y_\alpha} - \frac{\partial}
{\partial y_\beta} \right)\Lambda\right)^{2} \right]
\end{eqnarray}
where
\begin{equation}
\Lambda =
\sum_{\mu < \nu} ln \frac{\theta_{1}^{\ast}\left(z_{\mu}-z_{\nu} \mid\tau
\right)}
{\theta_{1}\left(z_{\mu}-z_{\nu} \mid\tau \right)}
\end{equation}

$H^{rel}$ can be understood as the generalization of the Hamiltonian
for the relative coordinates on the plane.  It is well known that the
$\theta_{1}\left(z\mid\tau \right)$ are the torus analogs of
$z=x+iy$ on the plane \cite{gaume}, and therefore the expression
$ln \frac{\theta_{1}^{\ast}\left(
z_{k}-z_{l} \mid\tau \right)}{\theta_{1}\left(z_{k}-z_{l} \mid\tau \right)}$
correponds to $ln \frac{z^\ast}{z}$ on the plane.  The Hamiltonian is
mapped to the free Hamiltonian by an obvious
transformation analogous to the one on the plane.  As far as $H^{rel}$ is
concerned everything is analogous to the case on the plane.

The center of mass Hamiltonian,
on the other hand, is quite a different object and there is no simple
analogy between it and the corresponding Hamiltonian on the plane.  On the
plane the center of mass Hamiltonian is explicitly free and knows nothing
about the flux tubes attached to the particles.  On the torus, however, the
Hamiltonian is not free, but, as I will show, there is a transformation
which takes it to a free form.  For such a transformation to exist it will
turn out to be necessary that the $\theta_i$ be quantized.  This crucial
difference will be responsible for the component structure of the anyon wave
functions on the torus.  In fact, I will show that if the $\theta_i$ are
not quantized
the picture of anyons as interacting Aharonov-Bohm tubes breaks down.

Since I am only interested in revealing the component structure of
anyon wave functions and not in finding exact solutions for the
entire Hamiltonian,
I will restrict my attention to the center of mass Hamiltonian in the
following.  In the previous section I discussed the non-trivial behavior
of the gauge field sector under the transformations $\theta_{i} \rightarrow
\theta_{i} + 2\pi$.  In particular it was shown that associated with each
physical value of $\exp i\theta_1$ there were $q$ linearly independent
Hilbert space rays any of which could be reached from any other by the
action of $T_1$ an appropriate number of times.
The second quantized Hamiltonian is invariant under the combined
transformations $T_{j}:\theta_{j} \rightarrow \theta_{j} + 2\pi$ and
$\psi \rightarrow \exp \left( \frac{i2\pi x_j}{L_j} \right) \psi$.  This
invariance is reflected in the first quantized theory by the set of conditions:
\begin{eqnarray}
T_{1}\phi\left(X,Y,\theta\right) = e^{i\gamma}\exp\left(i\frac{2\pi X}
{L_1} \right) \phi\left(X,Y,\theta \right) \\
T_{2}\phi\left(X,Y,\theta \right) = e^{i\beta}\exp\left(
i\frac{2\pi Y}{L_2} \right) \phi\left(X,Y,\theta \right)
\end{eqnarray}
Due to the
non-commutativity of the operators which translate the $\theta_i$
it is not possible to simultaneously impose the above conditions.
Instead, the most general conditions one may impose consistent with
the commutation relations of the $T_i$ are:
\begin{eqnarray}
T_{1}\phi_{l}\left(X,Y,\theta\right) & = & e^{i\gamma}
\exp\left(i\frac{2\pi XN_A}{L_1} \right) \phi_{l-1}\left(X,Y,\theta \right) \\
T_{2}\phi_{l}\left(X,Y,\theta\right) & = & e^{-i\beta -
i2\pi\kappa l}
\exp\left(i\frac{2\pi YN_A}{L_2} \right) \phi_{l}\left(X,Y,\theta \right)
\end{eqnarray}
The second condition states that
\begin{eqnarray}
\exp -i\left(\kappa\theta + 2\pi N_{A}Y/L_{2}\right)
\phi_{l}\left(X,Y,\theta \right) = \exp -i\left(\beta + 2\pi\kappa l \right)
\phi_{l}\left(X,Y,\theta \right)
\end{eqnarray}
The first condition requires that
\begin{eqnarray}
\phi_{l}\left(X,Y,\theta + 2\pi \right) = e^{i\gamma}
\exp\left(i\frac{2\pi XN_A}{L_1} \right) \phi_{l-1}\left(X,Y,\theta \right)
\end{eqnarray}

Turning now to the Hamiltonian, note that:
\begin{equation}
\left[\left(\frac{\partial}{\partial Y} - \frac{2\pi N_A}{\kappa L_2}
\frac{\partial}{\partial\theta}\right), \left(\frac{N_A}{L_1}\theta +
\frac{2\pi {N_A}^{2}}{\kappa L_{1}L_{2}} Y \right)\right] = 0
\end{equation}
This tells us that the operators $\partial_X$, $\left(\frac{N_A}{L_1}\theta +
\frac{2\pi {N_A}^{2}}{\kappa L_{1}L_{2}} Y \right)$, and
$\left({\partial_Y} - \frac{2\pi N_A}{\kappa L_2}
\frac{\partial}{\partial\theta}\right)$ can be diagonalized simultaneously.
The situation is complicated by the fact
that the Hamiltonian is not periodic on the torus (because of the presence
of a non-zero flux) making it necessary to impose boundary conditions which do
not
respect the commutativity of these operators.  In particular, one has to
sum over eigenstates of the Hamiltonian which carry distinct
eigenvalues
of the operator $P_X = -i\partial_X$ which makes it impossible to
diagonalize  $P_X$
(this is eloquently explained in the appendix in \cite{manton}).
What is important, though,
is that one can immediately write down the
eigenstates of the Hamiltonian from which one can construct solutions
obeying the correct boundary conditions.
\begin{eqnarray}
\chi_{l}^{{\vec k}{\vec m}}\left( X,Y,\theta\right) =
e^{i\left(2\pi k_{1}X/L_{1} \right)}
e^{i\left(2\pi k_{2}Y/L_{2} \right)}
e^{i\left(\frac{\gamma}{2\pi}+m_{2} \right)\theta }
\delta
\left(\kappa\theta + \frac{2\pi YN_A}{L_2} - \beta -2\pi\kappa l  - 2\pi
m_{1}\right)
\end{eqnarray}
are eigenstates of the Hamiltonian but do not satisfy the correct
quasi-periodic boundary conditions in $Y$ and $\theta$.
That is the wave function does not satisfy equation (38)
and the quasi-periodicity condition in $Y$:
\begin{equation}
\phi\left(X,Y+L_{2},\theta \right) =
\exp \left( \frac{i2\pi N_{A}^{2}X}{\kappa L_1} \right)
\phi\left( X,Y,\theta \right).
\end{equation}
The argument of the delta function in $\chi$ has been chosen so as to satisfy
equation (37).
It is a simple
excercise to show that the correct combination of these eigenstates is:
\begin{eqnarray}
& &\Phi_{ln}^{{\vec k}{\vec m}}\left( X , Y, \theta \right) =
\sum_{j=-\infty}^{\infty}
\chi^{k_{1}+{N_A}l+\frac{N_A}{\kappa} \left(pj+n\right),k_{2},{m_2},
\left(jp+n+m_{1} \right)}
\nonumber \\
& = &e^{i\frac{2\pi Y}{L_{2}}\left(k_{2} - \frac{N_A}{\kappa}
\left(m_{2} +\gamma/2\pi \right)\right)}
e^{i\left(\gamma/2\pi + m_{2}\right)
\left(\theta + 2\pi N_{A}Y/\kappa L_{2}\right)}
 \nonumber \\
& &\sum_{j=-\infty}^{\infty}  e^{\frac{i2\pi X}{L_1}\left( k_{1}+lN_{A} +
\frac{N_A}{\kappa} \left( pj+n \right) \right)}
\delta
\left(\kappa\theta + \frac{2\pi YN_A}{L_2} - \beta -2\pi\kappa l  - 2\pi
\left(jp +n+m_{1} \right) \right)
\end{eqnarray}

The index $n=0,...,p-1$ is a degeneracy index.
Note that these wave functions are non-zero for only a discrete set of
values of $\kappa\theta + 2\pi N_{A}Y/L_2$.
They are eigenstates of the Hamiltonian with eigenvalues:
\begin{eqnarray}
E_{{\vec k}{\vec m}} = \frac{1}{2mN_A}\left(\left( \frac{2\pi}{L_1}k_{1}
-\frac{N_A}{\kappa L_1}\left(\beta+2\pi m_{1} \right)
 \right)^2 +\left( \frac{2\pi}{L_2}k_{2}
-\frac{N_A}{\kappa L_2}\left(\gamma+2\pi m_{2} \right)
 \right)^2 \right)
\end{eqnarray}
The energy is independent of the gauge index $l$ as one would like, and
most importantly it is of the free particle form.  Interestingly,
$\beta$ and $\gamma$, the global phases allowed in
the most general boundary conditions for $\theta$, shift the energy spectrum.

\section{Single Valued Wave Functions on the Torus}
In this section I will show that it is possible to define single valued
wave functions on the torus.  It was shown above that the wave function
must pick up a gauge transformation after translation around at least one
cycle.  Therefore, it is necessary to specify what one means by single valued
wave functions in this context.  The wave function should really be thought
of as a section on the $U\left(1\right)$-bundle over the torus.  So a well
defined wave function on the torus should transform as a section which is
completely specified once one has picked a set of transition functions.
Alternatively, one may pick the translation operators to be the following:
\begin{equation}
t_{\alpha}\left(L_{i}{\hat e}_{i} \right) = \exp\left(-i\Lambda_{i}\right)
\exp\left(L_{i}\partial /\partial x_{\alpha i} \right)
\end{equation}
under the action of which the wave function should be single valued.  The
group of such translations is known in the condensed matter literature
as the magnetic translation group.  It is important to note that in the
case considered here the generators of magnetic translations $P_{\alpha i} =
\left({\partial}/{\partial x_{\alpha i}} - i\Lambda_{i}/L_{i} \right)$ do
not commute with the Hamiltonian, but finite translations by lattice vectors do
commute with the Hamiltonian.  The situation is, therefore, somewhat different
from the case of a constant magnetic field where this group usually appears.

This group of translations gives a precise way of formulating what one means
by single valued wave functions on the torus: that the wave
functions should be single valued under the group of magnetic translations.
The magnetic translation
group has been studied extensively in the condensed matter literature.
Haldane has studied this group for the many body case \cite{hal}.
I will follow Haldane's analysis, providing slightly more detail where
necessary, and show that
there is no obstruction to defining single valued wave functions on the torus.
	First, define the following operators:
\begin{eqnarray}
t_{\alpha} \left( {\vec a}_{i}\right) = \exp {\vec P}_{\alpha}\cdot{\vec a}
\end{eqnarray}
and note that
\begin{eqnarray}
t_{\alpha}\left({\vec a} \right)t_{\alpha}\left({\vec b} \right)
=t_{\alpha}\left({\vec b} \right)t_{\alpha}\left({\vec a}\right)
\exp\left(-i\frac{2\pi N_A}{\kappa L_{1}L_{2}}\left( {\vec a}\times {\vec b}
\right) \right)
\end{eqnarray}
When ${\vec a}$ and ${\vec b}$ are restricted to be of the form
${\vec L}_{mn} = m{\vec L}_{1} + n{\vec L}_{2}$, where
$m$,$n$ are integers, the translation operators
above commute among themselves and with the Hamiltonian and may be
simultaneously diagonalized.  To see how one
diagonalizes these operators it is convenient to factorize them into
a center of mass translation $T$ and a relative translation $\tilde{t}$
as follows:
\begin{equation}
t_{\alpha} \left( {\vec L}_{mn} \right) = T\left( \frac{{\vec L}_{mn}}{N_A}
\right)\tilde{t}_{\alpha}\left( {\vec L}_{mn} \right)
\end{equation}
where
\begin{eqnarray}
& &T\left( {\vec a} \right) = \prod_{\alpha} t_{\alpha}\left( {\vec a} \right)
\nonumber \\
& &\tilde{t}_{\alpha} \left( {\vec a} \right) =
T\left(- \frac{{\vec a}}{N_A}\right)
t_{\alpha}\left( \frac{{\vec a}}{N_A} \right)
\end{eqnarray}
The center of mass translation operator $T\left( {\vec a} \right)$ and
the relative translation operators
\{ $\tilde{t}_{\alpha}\left( {\vec b} \right)$ \}
commute for arbitrary arguments.

{}From the above commutation relations it is readily established that:
\begin{eqnarray}
& &\tilde{t}_{\alpha}\left({\vec L}_{1} \right)\tilde{t}_{\beta}
\left({\vec L}_{2} \right)
=\tilde{t}_{\beta}\left({\vec L}_{2} \right)\tilde{t}_{\alpha}
\left({\vec L}_{1}\right)
\exp i\frac{2\pi}{\kappa} \nonumber \\
& &T\left(\frac{{\vec L}_{1}}{N_A}\right)
T\left(\frac{{\vec L}_{2}}{N_A}\right)
 = T\left(\frac{{\vec L}_{2}}{N_A}\right) T\left(\frac{{\vec L}_{1}}{N_A}
\right) \exp -i\frac{2\pi}{\kappa}
\end{eqnarray}
These commutation relations imply that it is not possible to write down
a single valued wave function of the form $\psi^{cm} \otimes \psi^{rel}$.
However, it is still possible to write a single valued wave function
of the form $\sum_{i} \psi^{cm}_{i} \otimes \psi^{rel}_{i}$.  To establish
this result I will now diagonalize a maximal subset of translation operators in
each sector and find the single valued combination.

The largest commuting set of center of mass translation operators is,
$T\left(p\frac{{\vec L}_{2}}{N_A}\right)$,
$T\left(\frac{{\vec L}_{1}}{N_A}\right)$.  On the set of functions I
defined in the previous section their action is:
\begin{eqnarray}
T\left(\frac{{\vec L}_{1}}{N_A}\right) \phi^{{\vec k}{\vec m}}_{ln}
& = & e^{i2\pi k_{1}/N_A} e^{i2\pi n/\kappa} \phi^{{\vec k}{\vec m}}_{ln} \\
\nonumber \\
T\left(\frac{{\vec L}_{2}}{N_A}\right)\phi^{{\vec k}{\vec m}}_{ln}
& = & e^{i2\pi k_{2}/N_A} \phi^{{\vec k}{\vec m}}_{l,n-1} \nonumber \\
T\left(p\frac{{\vec L}_{2}}{N_A}\right)\phi^{{\vec k}{\vec m}}_{ln}
& = & e^{i2\pi pk_{2}/N_A} \phi^{{\vec k}{\vec m}}_{ln}
\end{eqnarray}
Similarly for the relative coordinate wave functions the maximal commuting set
of magnetic translation operators are
$\tilde{t}_{\alpha}\left({\vec L}_{1} \right)$,
$\tilde{t}_{\alpha}\left(p{\vec L}_{2} \right)$.  Now let
$\psi^{{\vec k}}_{\zeta s}$
stand for an eigenstate of the relative coordinate Hamiltonian satisfying
the following:
\begin{eqnarray}
\tilde{t}_{\alpha}\left({\vec L}_{1} \right)\psi^{{\vec k}}_{\zeta n}
& = & e^{-i2\pi k_{1}/N_A} e^{-i2\pi n/\kappa} \psi^{{\vec k}}_{\zeta n}
\nonumber \\
\tilde{t}_{\alpha}\left(p{\vec L}_{2} \right)\psi^{{\vec k}}_{\zeta n}
& = & e^{-i2\pi pk_{2}/N_A} \psi^{{\vec k}}_{\zeta n}
\end{eqnarray}
Then by the commutation relations:
\begin{eqnarray}
\tilde{t}_{\alpha}\left({\vec L}_{2} \right)\psi^{{\vec k}}_{\zeta n}
= e^{-i2\pi k_{2}/N_A} \psi^{{\vec k}}_{\zeta, n-1}
\end{eqnarray}
The index $\zeta$ keeps track of any other quantum numbers which the
relative wave function may carry.

The above classification of the center of mass and relative wave functions
in terms of magnetic translation eigenstates allows one to immediately
write down the total single valued wave function:
\begin{eqnarray}
\Psi^{{\vec k}{\vec m}}_{l \zeta} =\sum_{n=1}^{p} \phi^{{\vec k}{\vec m}}_{ln}
 \psi^{{\vec k}}_{\zeta n}
\end{eqnarray}
Notice that I have {\em not} shown that for any vector ${\vec k}$ there exists
a state $\psi^{{\vec k}}_{\zeta n}$ satisfying the above conditions.
One should really think of the quantum numbers ${\vec k}$ as being provided
by the relative wave function which (possibly) restrict the allowed values
of ${\vec k}$ appearing in the center of mass wave function to keep the
total wave function single valued.

It is important to note that there is no component structure when
the unconstrained part of the Chern-Simons gauge field is present in
the Hamiltonian.  This is
in contrast to what is claimed in ref \cite{ho} where a different
Schr\"{o}dinger wave function is defined.

\section{Component Structure of free Anyon Wave Functions}
It has long been known that for anyons on the plane one can adopt
one of two pictures \cite{lm}.
One works either with explicitly "anyonic"
(multivalued) wave functions or with single valued wavefunctions but with
Aharonov-Bohm flux tubes attached to the particles.  The equivalence of
the two pictures is established by constructing a unitary transformation which
maps the single valued wave functions to anyonic wave functions
satisfying the Schr\"{o}dinger equation with respect to a Hamiltonian without
flux tubes.

	The Chern-Simons term is often added as a convenient way
of attaching flux tubes to the particles.  Indeed, on the plane, the constraint
determines the gauge field completely and the C-S term's sole purpose is to
attach these flux tubes to the particles.
As is well known and shown above,
on the torus the C-S term does more than just attach flux tubes: it also
quantizes the flux quanta flowing through the holes of the torus.
It may appear that the intuitive picture of anyons as Aharonov-Bohm
tubes is a full description of anyons even on the torus and therefore
that the presence of a Chern-Simons term yields a non-minimal description of
anyons.  I would like to show that this is
not the case, that one {\em must} quantize the $\theta_i$ for the
single valued wave functions to be related by a unitary transformation to
the anyonic wave functions.

The negative result that when the $\theta_i$ are {\em not} quantized
there is no unitary transformation relating the single valued wave functions
to free anyon wave functions follows from a straightforward argument.
First suppose that
the constraint $f_{12} = -\frac{2\pi}{\kappa}J_0$ is given but that it is not
generated by a Chern-Simons term.  One would proceed as before only
now, due to the absence of the C-S term, the $\theta_i$ are not quantized but
are c-numbers.  The center of mass Hamiltonian is then given by
\begin{eqnarray}
H^{cm} = -\frac{1}{2mN_A} \left[ \left(\frac{\partial}{\partial X}
-i\frac{N_{A}\theta_{1}}{L_1} -i \frac{2\pi {N_A}^{2}}{\kappa L_{1}L_{2}}Y
\right)^{2} + \left( \frac{\partial}{\partial Y}-i\frac{N_{A}\theta_{2}}{L_2}
\right)^{2} \right]
\end{eqnarray}
But this is just the Hamiltonian for a particle of mass $mN_A$
in a constant magnetic field
of strength $b = -\frac{2\pi {N_A}^2}{\kappa L_{1}L_{2}}$.  There is no unitary
transformation which takes this Hamiltonian to a free form.  Moreover, the
energy spectrum is of the form $E= \omega\left(n+\frac{1}{2}\right)$
which is not at
all of the free form established above.  Therefore, pure
Aharonov-Bohm flux tubes is not a description of anyons on the torus.

	The positive result that there is a unitary equivalence between
free anyons and particles interacting with a Chern-Simons gauge field
on the torus remains to be established.  To construct the transformation
I remind the reader that
\begin{equation}
\left[\left(\frac{\partial}{\partial Y} - \frac{2\pi N_A}{\kappa L_2}
\frac{\partial}{\partial\theta}\right), \left(\frac{N_A}{L_1}\theta +
\frac{2\pi {N_A}^{2}}{\kappa L_{1}L_{2}} Y \right)\right] = 0
\end{equation}
Define a new center of mass wave function
$\tilde{\phi}_{ln}^{{\vec k}{\vec m}}$ by:
\begin{eqnarray}
\phi_{ln}^{{\vec k}{\vec m}} = \exp \left[ i\frac{N_{A}X}{L_1}
\left(\theta + \frac{2\pi N_A}{\kappa L_2}Y \right)\right]
\tilde{\phi}_{ln}^{{\vec k}{\vec m}}
\end{eqnarray}
$\tilde{\phi}_{ln}^{{\vec k}{\vec m}}$ satisfies the
Schr\"{o}dinger equation for a free particle of mass $mN_A$.  The
"gauge" transformed center of mass Hamiltonian depends only on the conjugate
momenta $-i\partial_{X}$ and
$-i\partial_{u} = -i\partial_{Y} + i\frac{2\pi N_A}{\kappa L_2}
\partial_{\theta}$ but not on the coordinates X and u=Y;
neither does it depend on the coordinate $v=\theta + \frac{2\pi N_A}{\kappa
L_2}Y$ nor its conjugate momentum $-i\partial_{v}=-i\partial_\theta$.
Hence the Hamiltonian
is free in $X$ and $u$ and is independent of $v$ and $-i\partial_v$.
{}From the form of the transformed Hamiltonian one would conclude that
$v$ is a constrained variable and its Hamiltonian evolution is trivial,
allowing arbitrary functions of $v$ to be perfectly good wave functions.
However, this is not the case, since the complicated boundary conditions that
the wave function must satisfy restrict its form considerably as I will
show by looking at the explicit form of the $\tilde{\phi}$.

$H^{rel}$ is brought to free form by writing
\begin{equation}
\psi^{{\vec k}}_{\zeta n} = \prod_{\alpha < \beta}
\left(\frac{\theta_{1}^{\ast}\left(z_{\alpha}-z_{\beta} \mid\tau \right)}
{\theta_{1}\left(z_{\alpha}-z_{\beta} \mid\tau \right)}
\right)^{\frac{1}{2\kappa}}
\tilde{\psi}^{{\vec k}}_{\zeta n}
\end{equation}
The wave function $\tilde{\psi}^{{\vec k}}_{\zeta n}$ solves the
Schr\"{o}dinger equation for
the free Hamiltonian in the relative coordinates.

Note that both gauge transformations are "singular".
For the gauge transformation on the relative coordinates it is well known
that there is no smooth well defined extrapolation to the points where
the particle coordinates coincide (the "origins" in the relative coordinates).
The center of mass gauge transformation (57) is singular in the sense that
if one embeds the torus in a higher dimensional space (e.g. in $R^3$)
there is no smooth well defined extrapolation to the entire space.  This
just means that at some point there is a source for the flux and there
is no way of gauging the source away.  But neither one of these singularities
is particularly troublesome since in the first case the wave function
vanishes identically whenever coordinates for any two particles coincide,
and in the
second one is always working on the surface of the embedded solid torus.

The transformed total wave function is

\begin{equation}
\Psi^{{\vec k}{\vec m}}_{l \zeta} =  \exp \left[ i\frac{N_{A}X}{L_1}
\left(\theta + \frac{2\pi N_A}{\kappa L_2}Y \right)\right]
 \prod_{\alpha < \beta}
\left(\frac{\theta_{1}^{\ast}\left(z_{\alpha}-z_{\beta} \mid\tau \right)}
{\theta_{1}\left(z_{\alpha}-z_{\beta} \mid\tau
\right)}\right)^{\frac{1}{2\kappa}}
\tilde{\Psi}^{{\vec k}}_{l \zeta 0}
\end{equation}
The index $0$ is included in anticipation of the component structure
to be revealed soon.
In terms of the center of mass and relative coordinates the total free
wave function is given by:
\begin{equation}
\tilde{\Psi}^{{\vec k}{\vec m}}_{l \zeta 0} = \sum_{n=1}^{p}
\tilde{\phi}^{{\vec k}{\vec m}}_{l n} \tilde{\psi}^{{\vec k}}_{\zeta n}
\end{equation}

Now I turn to the behavior of this wave function under translation by a lattice
vector.  First, define the following set of wave functions:
\begin{equation}
\tilde{\Psi}^{{\vec k}{\vec m}}_{l \zeta j} =
\sum_{n=1}^{p} e^{-i\frac{2\pi nj}{\kappa}}
\tilde{\phi}^{{\vec k}{\vec m}}_{ln} \tilde{\psi}^{{\vec k}}_{\zeta n}
\end{equation}
For $j=0$ the original wave function is recovered.
When one translates particle $\alpha$ by a lattice vector in the $x$-direction,
$x_{\alpha} \rightarrow x_{\alpha} + L_{1}$, while keeping all others
fixed, the wave function changes in the following way:
\begin{equation}
\tilde{\Psi}^{{\vec k}{\vec m}}_{l \zeta j} \left( x_{\alpha}+ L_{1} \right)
= \left( e^{-i\pi/\kappa} \right)^{N_{A} - 1} e^{-i\left(\beta/\kappa +
2\pi m_{1}/\kappa \right)}
\tilde{\Psi}^{{\vec k}{\vec m}}_{l \zeta, j+1} \left( x_{\alpha} \right)
\end{equation}
Thus $p$ translations of any particle in the $x$ direction generates the
entire set of linearly independent wave functions defined above.
When one translates particle $\alpha$ by a lattice vector in the $y$-direction
the above wave functions transform as follows:
\begin{equation}
\tilde{\Psi}^{{\vec k}{\vec m}}_{l \zeta j} \left( y_{\alpha}+ L_{2} \right)
= \left( e^{i\pi/\kappa} \right)^{N_{A} - 1} e^{-i2\pi j/\kappa}
\tilde{\Psi}^{{\vec k}{\vec m}}_{l \zeta, j} \left( y_{\alpha} \right)
\end{equation}
This shows that the transformed wave function has $p$-components.

One view of the component structure is provided by looking at the
singular gauge transformation on the center of mass wave function:
\begin{equation}
\exp \left[-i\frac{N_{A}X}{L_1}\left( \theta + \frac{2\pi N_{A}Y}{L_2} \right)
\right]
\end{equation}
Since the total wave function is not an eigenstate of
$\exp i\left(\theta +  \frac{2\pi N_{A}Y}{L_2} \right)$,
the wave function can not pick up a total phase under translations
$X \rightarrow X+L_{1}/N_{A}$ but is instead transformed
to an entirely different wave function.
This is a reflection of the fact that the translation group was
represented projectively and to
make the action of the translation group on the total wave function
single component and single valued
one had to sum over distinct eigenstates of $\exp iv$.
{}From this point of view the component structure arises due to the
fact that translation symmetry is realized projectively in the original
problem forcing one to abandon the $\psi^{cm}\otimes\psi^{rel}$
form of the wave functions in favor of single component and single valued
wave functions which do
not have a simple product form and are hence not eigenstates of $\exp iv$.

Another approach to understanding this phenomenon is provided by
working in a different basis.
\begin{equation}
\tilde{\Phi}^{{\vec k}{\vec m}}_{l \zeta n} =
\tilde{\phi}^{{\vec k}{\vec m}}_{l n} \tilde{\psi}^{{\vec k}}_{\zeta n}
\end{equation}
These wave functions are eigenstates of translations in the $x$-direction
but not in the $y$-direction.  They can be written in a factorized form:
\begin{eqnarray}
\tilde{\Phi}^{{\vec k}{\vec m}}_{l \zeta n}
& = & \left[ e^{i\frac{2\pi Y}{L_{2}}\left( k_{2}-N_{A}\left(\gamma + 2\pi
m_{2}\right) /\kappa \right)}
e^{\frac{i2\pi X}{L_1}\left( k_{1}-N_{A}\left(\beta + 2\pi m_{1}\right) /\kappa
+ \frac{N_A}{\kappa}n \right) } \right]
e^{-i\frac{2\pi N_{A}Xn}{\kappa L_1}} \nonumber \\
&  & \left[ \sum_{j=-\infty}^{\infty}
\exp {i\left(\gamma/2\pi + m_{2}\right)\left(\beta/\kappa +2\pi l +
2\pi /\kappa \left(jp +n+m_{1} \right)\right)} \right. \nonumber \\
&  & \left. \delta
\left(\kappa\theta + \frac{2\pi YN_A}{L_2} - \beta -2\pi\kappa l  - 2\pi
\left(jp +n+m_{1} \right) \right) \right] \tilde{\psi}^{\vec k}_{\zeta n}
\end{eqnarray}
The second factor in square brackets depends only on
$v=\theta + 2\pi N_{A}Y/\kappa L_{2}$, and is not periodic under translations
in the $y$ direction.  In fact, it is this part of the wave function which
is responsible for the component structure.  Wave functions in $v$ have
trivial Hamiltonian evolution (since the transformed
Hamiltonian is independent of $v$ and its conjugate momentum).
The wave functions in $v$ written above are exactly those
found by Polychronakos \cite{poly}
for pure Chern-Simons with $\theta$ replaced by $v$ and generalized to
cases where $q \neq 1$.  Usually a variable with respect to which
the Hamiltonian vanishes is allowed to have arbitrary wave functions.
However, as Polychronakos showed for the case of pure Chern-Simons,
when one has a compact phase space the Hilbert space is
finite dimensional.  This is exactly what has happened here
\footnote{Taking $u= Y$
gives $-i\partial_{v}=-i\partial_{\theta}=\frac{\kappa}{2\pi}\theta_2$.
Clearly $v$ and $p_v$ generate a compact phase space
($\theta_2$ and $\theta_{2}+2\pi$
are identified)}.
The Hilbert space of wave functions in $v$ is $p$-dimensional.
Thus the component structure really comes about
due to the appearance of a variable which generates a finite ($p$) dimensional
Hilbert space of wave functions which have trivial Hamiltonian evolution.
This variable is a combination of particle and
gauge degrees of freedom of such a form that the wave functions turn out
not to be periodic under lattice translations.  So in this second view
the component structure arises due to a peculiar mixing of gauge field
and particle degrees of freedom which do not allow one to have a single
component wave function.  These wave functions are completely analogous
to the ones which appear for the particle vacuum (the pure Chern-Simons case).

{}From the point of view of Chern-Simons theory what is surprising is not
that the anyon wave functions have multiple components but that they
only have a finite number of components.  The reason is that one starts out
with $2N_{A} +1$ coordinates but in the end one has only $2N_A$
coordinates with non-trivial Hamiltonian evolution.
One would expect that the
degree of freedom which has been "gauged" away
should generate an infinite degeneracy of which some
infinite set is related by lattice translations.  Instead one finds that this
degree of freedom generates only a finite dimensional Hilbert space!
This is a very surprising result indeed.

I close with a comment.  The gauge transformation used above to get
anyon wave functions is not the
most general.  One may always supplement the transformation with an extra
factor
$\exp iav$ where a is arbitrary.  Since $a$ and $a+1$ are related by
an $x$-translation both transformations project to members of the same
$p$ dimensional space of anyon wave functions.  Therefore one may
always restrict $0 \leq a<1$, this gives us the entire one parameter family
of possible anyon wave functions.  The parameter $a$ just shifts the
eigenvalues of $y$-translation by a constant amount.  The correspondence
between free anyons to bosons coupled to a Chern-Simons gauge field is
therefore many to one.

\section{Acknowledgments}
I am indebted to T. H. Hansson, A. Karlhede and M. Ljungberg for discussions
while this work was in progress and for reading the manuscript critically.
I am also indebted to Marcelo Camperi for introducing me to anyons
and Chern-Simons theories and bringing ref \cite{lee} to my attention.
I thank F. Bastianelli and E. Westerberg for stimulating discussions.

\end{document}